\documentclass[prd, twocolumn, superscriptaddress]{revtex4}

\usepackage{epsfig}
\usepackage{graphicx}
\usepackage{amsmath}
\usepackage{amsfonts}
\usepackage{epstopdf}

\def\slashchar#1{\setbox0=\hbox{$#1$}     		
   \dimen0=\wd0                                 	
   \setbox1=\hbox{/} \dimen1=\wd1               	
   \ifdim\dimen0>\dimen1                        	
      \rlap{\hbox to \dimen0{\hfil/\hfil}}      	
      #1                                        	
   \else                                        	
      \rlap{\hbox to \dimen1{\hfil$#1$\hfil}}   	
      /                                         	
   \fi}

\renewcommand{\vec}{\boldsymbol}

\begin{document}

\title{Chiral Magnetic conductivity}

\author{Dmitri E. Kharzeev}
\email{kharzeev@bnl.gov}
\affiliation{Department of Physics,
Brookhaven National Laboratory, Upton NY 11973, USA}

\author{Harmen J. Warringa}
\email{warringa@th.physik.uni-frankfurt.de}
\affiliation{Institut f\"ur Theoretische Physik, Goethe-Universit\"at
  Frankfurt, Max-von-Laue-Stra\ss e~1, 60438 Frankfurt am Main, Germany}

\date{\today}

\begin{abstract}
  Gluon field configurations with nonzero topological charge generate 
  chirality, inducing $\mathcal{P}$- and $\mathcal{CP}$-odd effects.
  When a magnetic field is applied to a system with nonzero chirality,
  an electromagnetic current is generated along the direction of the
  magnetic field.  The induced current is equal to the Chiral Magnetic
  conductivity times the magnetic field.  In this article we will
  compute the Chiral Magnetic conductivity of a high-temperature
  plasma for nonzero frequencies.  This allows us to discuss the
  effects of time-dependent magnetic fields, such as 
  produced in heavy ion collisions, on chirally asymmetric systems.
\end{abstract}

\maketitle

\section{Introduction}
Quantum chromodynamics (QCD) contains gauge field configurations which
carry topological charge \cite{BPST}.  These configurations
interpolate between the different vacua of the gluonic sector of QCD
\cite{CDG76} and induce interesting $\mathcal{P}$- and
$\mathcal{CP}$-odd effects \cite{H}. Neglecting the masses of quarks
(as appropriate at high energies), it holds for each individual flavor
that \cite{Anomaly}
\begin{equation}
 \Delta N_5 \equiv \Delta(N_R - N_L) = - 2 Q,
 \label{eq:anomaly}
\end{equation}
where $\Delta N_5$ denotes the change in chirality ($N_5$) which is the
difference between the number of modes with right- and left-handed
chirality.  In the limit of zero quark mass $N_5$ is also equal to the
total number of particles plus antiparticles with right-handed
helicity minus the total number of particles plus antiparticles with
left-handed helicity. Right-handed helicity means that spin and
momentum are parallel, whereas left-handed helicity implies they are
opposite.

There exist different mechanisms to generate topological charge during
heavy ion collisions. One possible way is by longitudinal fields
created just after the collision \cite{KKV, LM06,Kharzeev:2006zm},
another due to QCD sphaleron transitions in the quark-gluon-plasma
\cite{MMS, BMR, SS02, SZ03}, and also plasma instabilities can lead to
generation of topological charge \cite{AM06}. Furthermore, it has been
argued that instanton "ladders" may describe a significant fraction of
multi-particle production at high energies
\cite{Kharzeev:2000ef,Shuryak:2000df,Nowak:2000de}. Finally,
metastable ${\cal P}-$ and ${\cal CP}-$ odd domains may exist in the
quark-gluon plasma close to the critical temperature
\cite{Kharzeev:1998kz}.  No matter what the precise mechanism behind
the generation of topological charge is, this leads to generation of
chirality as can be seen from Eq.~(\ref{eq:anomaly}).

Assuming that the $\theta$ angle vanishes there is no explicit
$\mathcal{P}$- and $\mathcal{CP}$-breaking in QCD. Hence positive and
negative topological charges are being generated with equal
probability. But because of fluctuations a finite amount of
topological charge can be generated in each event within a given
region of phase space determined by the experimental acceptance. Only
if one averages over many events the net produced topological charge
should vanish. Therefore in each event a difference between the total
number of right- and left-handed particles can be expected.

When two heavy ions collide with nonzero impact parameter, a
(electromagnetic) magnetic field of enormous magnitude is created in
the direction of angular momentum of the collision; it has been
evaluated in Refs \cite{KMW, SIT09} (see also \cite{MM96} for a
proposal to utilize electromagnetic fields in the search for
disoriented chiral condensates).  If a nonzero chirality
is present in such a situation, an electromagnetic current will be
induced in the direction of the magnetic field. This is the so-called
Chiral Magnetic Effect \cite{KMW, K06, KZ, FKW}; see recent
Ref.~\cite{BCLP09} for the first study of this effect in lattice QCD.
To understand this effect qualitatively let us imagine a situation in
which the $\mathcal{P}$-and $\mathcal{CP}$-odd processes made $N_5$
positive, so that we have an excess of quarks plus antiquarks with
right-handed helicity. In a background magnetic field, the quarks will
align their magnetic moments along the magnetic field. And assuming
the quarks can be treated as massless, the momenta of the quarks will
be aligned along the field as well. Consequently a quark with
left-handed helicity tends to move exactly in the opposite direction
to a quark with right-handed helicity. Since the magnetic moment is
proportional to the charge, an antiquark with right-handed helicity
will move exactly opposite to a quark with right-handed
helicity. Accordingly, in this case of positive $N_5$ an excess of
positive charge will move parallel to the magnetic field and an excess
of negative charge will go in the opposite direction. Thus an
electromagnetic current is generated along the magnetic field.

In a heavy ion collision this current leads to an excess of positive
charge on one side of the reaction plane (the plane in which the beam
axis and the impact parameter lies) and negative charge on the other;
the resulting charge asymmetry is also modulated by the radial flow
and the transport properties of the medium. This charge asymmetry can
be investigated experimentally using the observables proposed in
Ref.~\cite{V04}. Preliminary data of the STAR collaboration has been
presented in Refs.~\cite{IVS, VQM}. Implications of the Chiral
Magnetic Effect on astrophysical phenomena have recently been
discussed in Ref.\ \cite{CZ09}; another astrophysical implication can
be found in \cite{Andrianov:2009tp}.

A system of massless fermions with nonzero chirality can be described by
a chiral chemical potential $\mu_5$ which couples to the zero
component of the axial vector current in the Lagrangian. The induced
current in such situation can be written as $\vec j = \sigma_{\chi}
\vec B$, where $\sigma_{\chi}$ is the Chiral Magnetic
conductivity. For constant and homogeneous magnetic fields its value
is determined by the electromagnetic axial anomaly and for one
flavor and one color equal to
\cite{NM, HG, FKW} (see also \cite{MZ})
\begin{equation}
  \sigma_{\chi}(\omega = 0, \vec p = 0) \equiv \sigma_0 = 
 \frac{e^2}{2\pi^2} \mu_5,
\label{eq:chicon0}
\end{equation}
where $\omega$ and $\vec p$ denote frequency and momentum
respectively, and $e$ equals the unit charge.  For a finite number of
colors $N_c$ and flavors $f$ one has to multiply this result by $N_c
\sum_f q_f^2$ where $q_f$ denotes the charge of a quark in units of
$e$. The generation of currents due to the anomaly in background
fields or rotating systems is also discussed in related
contexts in Refs.~\cite{GW81, CH85, MZ, GMS09, SS09}.

For constant magnetic fields which are inhomogeneous in the plane
transverse to the field one finds that the total current $J$ along $B$
equals \cite{FKW},
\begin{equation}
  J = e 
\Bigl \lfloor
\frac{e \Phi}{2\pi} \Bigr \rfloor \frac{L_z \mu_5}{\pi},
\label{eq:curinhom}
\end{equation}
where $L_z$ is the length of the system in the $z$-direction
and the flux $\Phi$ is equal to the integral
of the magnetic field over the transverse plane,
\begin{equation}
\Phi =  \int \mathrm{d}^2 x\, B(x,y).
\end{equation}
The floor function $\lfloor x \rfloor$ is the largest integer smaller
than $x$. The quantity $\lfloor e \Phi/ (2\pi)\rfloor$ in
Eq.~(\ref{eq:curinhom}) is equal to the number of zero modes in the
magnetic field \cite{AC}.

To compute the current generated by a configuration of specific
topological charge, one should express $\mu_5$ in terms of the
chirality $N_5$. By using the anomaly relation one can then relate
$N_5$ to the topological charge. This is discussed in detail in
Ref.~\cite{FKW}.

The aim of this paper is to study how a system with constant nonzero
chirality responds to a time dependent magnetic field. This is
interesting for phenomenology since the magnetic field produced with
heavy ion collisions depends strongly on time. To obtain the induced
current in a time-dependent magnetic field, we will compute the Chiral
Magnetic conductivity for nonzero frequencies and nonzero momenta
using linear response theory. We will compute the leading order
conductivity and leave the inclusion of corrections due to photon and
or gluon exchange for future work.  In leading order the Chiral
Magnetic conductivity for an electromagnetic plasma and quark gluon
plasma are equal (up to a trivial factor of $N_c \sum_f q_f^2$).
Since we do not take into account higher order corrections, some of
our results for QCD will only be a good approximation in the limit of
very high temperatures where the strong coupling constant $\alpha_s$
is sufficiently small.

We will take the metric $g^{\mu \nu} = \mathrm{diag}(+, -, -, -)$.
The gamma matrices in the complete article satisfy $\{ \gamma^\mu,
\gamma^\nu \} = 2 g^{\mu \nu}$.  We will use the notation $p$ for both
the four-vector $p^\mu = (p_0, \vec p)$ and the length of a
three-vector $p = \vert \vec p \vert$. 

\section{Kubo formula for Chiral Magnetic Conductivity}
For small magnetic fields, the induced vector current can be found
using the Kubo formula.  This formula tells us that to first order in
the time-dependent perturbation, the induced vector current is equal
to retarded correlator of the vector current with the perturbation
evaluated in equilibrium. More explicitly, one finds that
\begin{equation}
  \langle j^\mu(x) \rangle 
= \int \mathrm{d}^4 x'\,
  \Pi^{\mu \nu}_{\mathrm{R}}(x, x') A_\nu(x'),
\end{equation}
where $j^\mu(x) = e\bar \psi(x) \gamma^\mu \psi(x)$ and
the retarded response function $\Pi_{\mathrm{R}}^{\mu \nu}$ is given by
\begin{equation}
\Pi^{\mu \nu}_{\mathrm{R}}(x, x')
=
i \langle
\left[ j^\mu(x), j^\nu( x') \right]
\rangle \theta(t-t')
.
\end{equation}
The equilibrium Hamiltonian is invariant under translations
in time and space, therefore we can use that
$\Pi^{\mu \nu}_{\mathrm{R}}(x, x')
=
\Pi^{\mu \nu}_{\mathrm{R}}(x-x')
$. 
Let us take a vector field of the following specific form $A_\nu(x) =
\tilde A_\nu(p) e^{-i p x}$.  The Kubo formula now becomes,
\begin{equation}  
\langle j^\mu(x) \rangle 
=
\tilde \Pi^{\mu \nu}_{\mathrm{R}}(p)
\tilde A_{\nu}(p) e^{-i px},
\label{eq:kubomo}
\end{equation}
where
\begin{equation}
\tilde \Pi^{\mu \nu}_{\mathrm{R}}(p)
= \int \mathrm{d}^4 x\,
e^{i p x} \Pi_{\mathrm{R}}^{\mu \nu}(x).
\end{equation}

In order to compute the Chiral Magnetic conductivity we will take a
time-dependent magnetic field pointing in the $z$-direction. 
Because of Faraday's law ($\vec \nabla \times \vec E = - \partial \vec B /
\partial t$), such time-dependent magnetic field comes always together
with a perpendicular electric field.  Let us choose a gauge such that
the only component of the vector field that is non-vanishing is $A_y$.
Then $B_z(x) = \partial_x A_y(x)$ so that $\tilde B_z(p) = i p^1
\tilde A^2(p)$. Using Eq.~(\ref{eq:kubomo}) we find that the induced
vector current in the direction of the magnetic field can now be
written as
\begin{equation}
\langle j_z(x) \rangle 
=
\sigma_{\chi}(p) \tilde B_z(p) e^{-ipx},
\end{equation}
where the Chiral Magnetic conductivity equals
\begin{equation}
\sigma_{\chi}(p) 
= 
\frac{1}{i p^1} \tilde \Pi^{2 3}_{\mathrm{R}}(p)
= \frac{1}{2 i p^i} \tilde \Pi^{jk}_{\mathrm{R}}(p) \epsilon^{ijk}.
\label{eq:chiconduc1}
\end{equation}
The right-hand side of the last equation is a result of rotational and
gauge invariance. We can now write the Chiral Magnetic conductivity in
the following way
\begin{equation}
\sigma_{\chi}(p) 
=  \frac{1}{i p^i} G^i_{\mathrm{R}}(p),
\label{eq:chiconduc2}
\end{equation}
with
\begin{equation}
G^i_{\mathrm{R}}(p) = 
\frac{1}{2} \epsilon^{ijk} \tilde \Pi_{\mathrm{R}}^{j k}(p).
\label{eq:gdef}
\end{equation}
The last two equations show that the Chiral Magnetic conductivity
follows directly from evaluating $G^i(p)$, which is the spatially
antisymmetric part of the off-diagonal retarded current-current
correlator, or equivalently the photon polarization tensor.

Another related quantity that follows from the off--diagonal part of
the photon polarization tensor is the Hall conductivity. The Hall
current is generated in the presence of a magnetic field that is
perpendicular to an electric field.  Unlike the Chiral Magnetic current,
the Hall current is in a direction perpendicular to the magnetic
field.  In order to obtain the Hall conductivity one usually computes
the photon polarization tensor in the presence of a homogeneous
background field.  The electric field is then treated as a
perturbation (see e.g. Ref.~\cite{LM09} for a recent calculation
using a holographic model of QCD).

In general, the Chiral Magnetic conductivity will be complex.  Let us
therefore write
\begin{equation}
\sigma_{\chi}(p) = \sigma_{\chi}'(p) + i \sigma_{\chi}''(p),
\end{equation}
where $\sigma_{\chi}'(p)$ and $\sigma_{\chi}''(p)$ are real functions given by
\begin{eqnarray}
\sigma_{\chi}'(p)  
&=&
\frac{1}{p^i} 
\mathrm{Im}\, G^i_{\mathrm{R}}(p),
\\
\sigma_{\chi}''(p)  
&=&
- \frac{1}{p^i} 
\mathrm{Re}\, G^i_{\mathrm{R}} (p) .
\label{eq:realimagconductivity}
\end{eqnarray}
For convenience we will write the zero momentum limit of the Chiral
Magnetic conductivity as follows
\begin{equation}
 \sigma_\chi(\omega) \equiv \lim_{\vec p \rightarrow 0}
 \sigma_\chi(p_0 = \omega, \vec p).
\end{equation}
The real part of the conductivity is an even function of $\omega$
while the imaginary part is odd, i.e. $\sigma_{\chi}'(\omega) =
\sigma_{\chi}'(-\omega)$ and $\sigma_{\chi}''(\omega) =
-\sigma_{\chi}''(-\omega)$.  If we apply a homogeneous magnetic field
$\vec B(t) = B_\omega \cos(\omega t) \hat{\vec z} $ the response $j(t) =
\langle j_z(t) \rangle$ will be
\begin{equation}
j(t) = \left[\sigma_\chi'(\omega) \cos(\omega t) + \sigma_\chi''(\omega) 
\sin(\omega t) \right] B_\omega 
\end{equation}
Hence the real part and imaginary part correspond to the in- and
out-phase response respectively.  To find the response to a general
time-dependent magnetic field $\vec B = B(t) \hat {\vec z}$ one
first has to compute the Fourier transform of the magnetic field
\begin{equation}
\tilde B(\omega) = \int^{\infty}_{-\infty}
\mathrm{d} t\, e^{i\omega t} B(t).
\end{equation}
The response will then be
\begin{equation}
 j(t) 
= 
\int^{\infty}_{0}
\frac{ \mathrm{d} \omega}{\pi}
\left[
\sigma'(\omega) \cos(\omega t) + 
\sigma''(\omega) \sin(\omega t) 
\right]
\tilde B(\omega).
\label{eq:response}
\end{equation}

The real and imaginary parts of the Chiral Magnetic conductivity
related to each other by the Kramers-Kroning relation
\begin{eqnarray}
\sigma_{\chi}'(\omega) &=& 
\frac{1}{\pi} \mathcal{P}
\int_{-\infty}^{\infty} \mathrm{d} q_0 \frac{\sigma_{\chi}''(q_0)}
{q_0 - \omega},
\label{eq:kk1}
\\
\sigma_{\chi}''(\omega) &=& 
- \frac{1}{\pi} \mathcal{P}
\int_{-\infty}^{\infty} \mathrm{d} q_0 
\frac{\sigma_{\chi}'(q_0)}{q_0 - \omega}.
\label{eq:kk2}
\end{eqnarray}

In the next section we will study the fermion propagator in the
presence of a chiral chemical potential. Then using this propagator we
will compute the retarded current-current correlator from which we can
derive the Chiral Magnetic conductivity.

\section{Fermion propagator}
The bare fermion propagator as a function of Euclidean momentum $Q$ in
the presence of a chiral chemical potential equals
\begin{equation}
S(Q) 
=
\frac{1}
{
 i \gamma^0 ( \tilde \omega_m 
- i \mu 
-
i \mu_5 \gamma^5
)
 -
\vec \gamma \cdot \vec q
},
\end{equation}
here $\tilde \omega_m = (2 m + 1)\pi T$ with $m \in \mathbb{Z}$ is a
fermionic Matsubara mode.  By computing the inverse, this propagator
can be written as
\begin{equation}
S(Q) 
=
 \mathcal{P_+} \frac{\slashchar{Q}_+}{Q_+^2}
+
\mathcal{P_-} \frac{\slashchar{Q}_-}{Q_-^2}, 
\label{eq:propmu5old}
\end{equation}
where we have defined $Q^\mu_\pm = (i\tilde \omega_\pm , \vec q)$ with
$\tilde \omega_\pm = \tilde \omega_m - i \mu_\pm$ and $\mu_\pm = \mu
\pm \mu_5$.  Accordingly $Q_{\pm\mu} = (i\tilde \omega_\pm, -\vec q)$
and $Q^2_\pm = -(\tilde \omega_\pm^2 + \vec q^2)$.  Furthermore
$\slashchar{Q}_\pm = Q_\pm^\mu \gamma_\mu$.  The right- ($+$) and
left-handed ($-$) chirality projection operators are given by
$\mathcal{P}_\pm = \tfrac{1}{2}(1 \pm \gamma_5)$. They satisfy
$\mathcal{P}_\mathrm{\pm}^2 = \mathcal{P}_\pm$ and $\mathcal{P}_+
\mathcal{P}_- = 0$. Let us now define
\begin{equation}
 \Delta_\pm(q_0, \vec q) = \frac{1}{q_0 \mp E_q},
\end{equation}
where $E_q = \vert \vec q \vert$. Furthermore we introduce $\hat{\vec
  q} = \vec q / \vert \vec q \vert$ and $\hat q^\mu_\pm \equiv (1, \pm
\hat {\vec q})$.  With these definitions the fermion propagator can be
rewritten as
\begin{equation}
S(Q) 
= \frac{1}{2} 
\sum_{s, t=\pm} 
\Delta_t(i \tilde \omega_s, \vec q)
\mathcal{P}_s
\gamma_\mu \hat q^\mu_t.
\label{eq:propmu5}
\end{equation}
As can be seen from Eq.~(\ref{eq:propmu5}) the propagator consists of
two parts describing the modes with right- ($s=+$) and left-handed
chirality ($s=-$). Because we took $m=0$ the opposite chiralities do
not mix.  The different values of $t$ correspond to particles (modes
with positive energy, $t=+$) and antiparticles (modes with negative
energy, $t=-$).  In the following section we will use this expression
for the propagator to compute the retarded current-current correlator.

\section{Computation of retarded correlator}
We will compute the retarded correlator $G_\mathrm{R}^i(p)$, defined
in Eq.~(\ref{eq:gdef}), using the imaginary time formalism of thermal
field theory.  The retarded correlator can be obtained from the
Euclidean correlator by analytic continuation in the following way
\begin{equation}
  G_{\mathrm{R}}^{i}(p_0, \vec p) =
    G^i_{\mathrm{E}}(\omega_n, \vec p) 
\vert_{i\omega_n \rightarrow p_0 + i \epsilon},
\end{equation}
where $\epsilon = 0^+$.

At very high temperatures, the gluons and quark masses can be
neglected to first approximation and the current-current correlator is
a convolution of two bare massless fermion propagators $S(Q)$ (see
e.g.\ Ref.\ \cite{Kapusta} for the correct expression). Using
Eq.~(\ref{eq:gdef}) we find
\begin{equation}
G_{\mathrm{E}}^{i}(P)
= \frac{e^2}{2 \beta}
\sum_{\tilde \omega_m} \int \frac{\mathrm{d}^3 q}{(2\pi)^3}
\epsilon^{ijk}
\mathrm{tr} 
\left[ \gamma^{k}
S(Q)
\gamma^{j}
S(P+Q)
\right].
\label{eq:photonpolarization}
\end{equation}

With use of Eq.~(\ref{eq:propmu5}) and the properties of the chirality
projection operators the integrand of
Eq.~(\ref{eq:photonpolarization}) can now be written as
\begin{multline}
\frac{1}{4} \sum_{s, t, u = \pm}
\epsilon^{ijk} \mathrm{tr} \left[
\gamma_\nu \gamma^k \gamma_\mu \gamma^j \mathcal{P}_s \right]
\times
\\
\Delta_t(i \tilde \omega_s, \vec q)
\Delta_{u}(i \tilde \omega_s + i \omega_n, \vec p + \vec q)
\hat q^\mu_t
(\widehat {p+ q} )^\nu_{u}.
\end{multline}
It can be seen from the last equation that the opposite chiralities
($s=\pm$) do not mix. As long as $m=0$, the Chiral Magnetic
conductivity is a sum of a contribution from purely right-handed modes
and purely left-handed modes.

We can now use that $\mathrm{tr}[\gamma^\mu \gamma^\nu \gamma^\rho
\gamma^\sigma \gamma^5] = - 4 i \epsilon^{\mu \nu \rho \sigma}$ where
$\epsilon^{\mu \nu \rho \sigma}$ is the complete antisymmetric tensor
with $\epsilon^{0 1 2 3} = 1$.  Then it follows that
\begin{equation}
\epsilon^{ijk} \mathrm{tr} \left[
\gamma_\nu \gamma^k \gamma_\mu \gamma^j \gamma^5 \right] a^\mu b^\nu
=
8 i \left( a^i b^0  - a^0 b^i \right)
.
\end{equation}
As a result we obtain
\begin{multline}
G_{\mathrm{E}}^{i}(P)
= 
 \frac{i e^2}{2 \beta}
\sum_{\tilde \omega_m} \int \frac{\mathrm{d}^3 q}{(2\pi)^3}
\sum_{s,t,u=\pm} 
s \left[
t \frac{q^i}{E_q}
-
u \frac{p^i + q^i}{E_{p+q}}
\right]
\times
\\
\Delta_t(i \tilde \omega_s, \vec q)
\Delta_{u}(i \tilde \omega_s + i \omega_n, \vec p + \vec q)
.
\label{eq:GE}
\end{multline}
We can now perform the sum over Matsubara frequencies.  Using that
$\omega_n = 2n \pi T$ is a bosonic Matsubara frequency one finds
\begin{multline}
 \frac{1}{\beta}
\sum_{\tilde \omega_m}
\Delta_t(i \tilde \omega_s, \vec q)
\Delta_{u}(i \tilde \omega_s + i \omega_n, \vec p + \vec q)
=
\\
 \frac{
t \tilde n(E_q - t \mu_s)
-
u \tilde n(E_{p+q} - u \mu_s)
+
\tfrac{1}{2} (u -t)
}
{i \omega_n + t E_q - u E_{p+q}}
,
\label{eq:matsusum}
\end{multline}
where $\tilde n(x) = [\exp(\beta x) + 1]^{-1}$ is the Fermi-Dirac
distribution function.  We can now perform the analytic continuation
in order to obtain the retarded correlator $G^i_\mathrm{R}(p)$, which
amounts to replacing $i \omega_n$ by $p_0 + i \epsilon$ in
Eq.~(\ref{eq:matsusum}). We obtain
\begin{multline}
G_{\mathrm{R}}^{i}(p)
= 
 \frac{i e^2}{2}
 \int \frac{\mathrm{d}^3 q}{(2\pi)^3}
\sum_{s,t,u=\pm} 
s 
\left[
t \frac{q^i}{E_q}
-
u \frac{p^i + q^i}{E_{p+q}}
\right]
\times
\\
\frac{
t \tilde n(E_q - t \mu_s)
-
u \tilde n(E_{p+q} - u \mu_s)
+ \tfrac{1}{2} (u - t)
}
{
p_0 + i \epsilon + t E_q - u E_{p+q}
}
.
\label{eq:GE1}
\end{multline}
Before we compute the integral over $q$ let us try to interpret
Eq.~(\ref{eq:GE1}). The retarded correlator $G^i_\mathrm{R}(p)$ has a
real part when $p_0 = u E_{p+q} - t E_q$. From
Eq.~(\ref{eq:realimagconductivity}) it can be seen that in that case
the Chiral Magnetic conductivity acquires an imaginary part. When $p_0
= u E_{p+q} - t E_q$ the virtual particles in the loop of the photon
polarization tensor go on the mass-shell, which by the optical theorem
corresponds to production or scattering of real particles in the
electromagnetic field.

If $u=-t$ this implies that particle antiparticle pairs are produced
from the electromagnetic field that oscillates with frequency $p_0$.
If $t=u$ particles ($u=t=1$) or antiparticles ($u=t=-1$) scatter from
the electromagnetic field and acquire or lose some momentum and
energy. Let us take a closer look at the pair production
process.

If $p_0 > 0$ the produced particles and antiparticles have energies
$E_{p+q}$ ($u=1$) and $E_{q}$ ($t=-1$) respectively. If the system we
consider consists mainly of particles, $\mu_s$ is positive. At zero
temperature all particle states up to the Fermi energy $\mu_s$ are
filled, so then it is impossible to produce particles with energy per
particle less than $\mu_s$ due to Pauli blocking. This is reflected in
the Fermi-Dirac distributions in the integrand of Eq.~(\ref{eq:GE1}),
for $u=t=-1$, $\mu_s > 0$, and $T=0$, the integrand vanishes if
$E_{p+q} < \mu_s$.  If $E_{p+q} > \mu_s$ there is no Pauli blocking,
in that case $E_q > \mu_s - p$. Hence at $T=0$, $G^i_\mathrm{R}(p)$
develops a real part and particle-antiparticle pairs with chirality
$s$ can be produced if $p_0 > 2 \mu_s - p$.  If both $E_{p+q}$ and
$E_q$ are larger than $\mu_s$ the produced pair does not feel the
influence of the nonzero chirality, so then the imaginary part of the
Chiral Magnetic conductivity should vanish as well. In that case $E_q
< \mu_s$ so that $E_{p+q} < \mu_s + p$ which gives $p_0 < 2 \mu_s +
p$.  So concluding we expect the Chiral Magnetic conductivity to have
an imaginary part for $2 \mu_s - p < p_0 < 2\mu_s + p$ due to pair
production.

Now let us continue the calculation of $G_{\mathrm{R}}^{i}(p)$.  The
last term in the integrand of Eq.~(\ref{eq:GE1}), vanishes after
summing over chiralities ($s$). It then follows that the function
$G_{\mathrm{R}}^{i}(p)$ is ultraviolet finite because the
high-momentum part of the integrand is exponentially suppressed by the
Fermi-Dirac distribution functions. We can therefore shift and reflect
the integration variable as follows $\vec q \rightarrow -\vec q - \vec
p$ in order to make both Fermi-Dirac distribution functions dependent
on $E_q$. Then after interchanging $t$ with $u$ and inversions $t
\rightarrow -t$ and $u \rightarrow -u$ we arrive at
\begin{multline}
G_{\mathrm{R}}^{i}(p)
= 
\frac{i e^2}{2}
 \int \frac{\mathrm{d}^3 q}{(2\pi)^3}
\sum_{s,t,u=\pm} 
s 
\left[
t \frac{q^i}{E_q}
+
u \frac{p^i + q^i}{E_{p+q}}
\right] \times
\\
\frac{
\tilde n(E_q - \mu_s) - 
\tilde n(E_q + \mu_s)
}{
p_0 + i \epsilon + t E_q + u E_{p+q}
}
.
\label{eq:GE2}
\end{multline}
Here we used that since $t=\pm$,
\begin{multline}
t \left[
\tilde n(E_q - t \mu_s) - 
\tilde n(E_q + t \mu_s) 
\right]
= \\
\tilde n(E_q -  \mu_s) - 
\tilde n(E_q + \mu_s).
\end{multline}
We can now perform the sum over $u$ which gives us
\begin{multline}
G_{\mathrm{R}}^{i}(p)
= 
i e^2
 \int \frac{\mathrm{d}^3 q}{(2\pi)^3}
\sum_{s,t =  \pm} 
s \left[
t \frac{p_0}{E_q} q^i
- p^i  \right]
\times
\\
\frac{
\tilde n(E_q - \mu_s) - 
\tilde n(E_q + \mu_s)
}{
(p_0 + i \epsilon + t E_q)^2 -E_{p+q}^2
}
.
\label{eq:GE3}
\end{multline}
Now it is possible to perform the angular integral. We use that the
fact that the integral has to be proportional to $p^i$. Hence we can
replace $q^i$ by $(\vec q \cdot \vec p) p^i / p^2$.  We find
after integrating over angles
\begin{multline}
G_{\mathrm{R}}^{i}(p)
= 
\frac{i e^2}{16 \pi^2} \frac{p^i}{p}
\frac{p^2 - p_0^2}{p^2}
 \int_0^\infty \mathrm{d} q
\,
f(q)
\times
\\
\sum_{t = \pm}
\left(
2 q
+ t p_0
\right)
\log \left[
\frac{ (p_0 + i \epsilon + t q)^2 - (q + p)^2}
{ (p_0 + i \epsilon + t q)^2 - (q - p)^2}
\right]
.
\label{eq:GER5}
\end{multline}
where
\begin{equation}
f(q) \equiv \sum_{s = \pm} 
s 
\left[ \tilde n(q - \mu_s) - 
\tilde n(q + \mu_s)
\right].
\end{equation}

We now have all results in order to obtain the Chiral Magnetic
conductivity which follows from combining Eq.~(\ref{eq:GER5}) with
Eq.~(\ref{eq:realimagconductivity}).  We will discuss the result in
the next section.

\section{Computation of chiral magnetic conductivity}
We will discuss the leading order contribution to the Chiral Magnetic
conductivity as a function of frequency and momentum. The zero
frequency and momentum value is constrained by the axial anomaly, so
loop corrections by gluons and/or photons will not alter this result
\cite{FKW}. However, loop corrections will change the conductivity at
nonzero frequencies.

In an electromagnetic plasma, the leading order result is a sensible
approximation since loop effects are of order $\alpha_{\mathrm{EM}}$
which is small. The results for an electromagnetic plasma are only
reliable if the temperature is larger than the mass of the electron,
since we have assumed massless particles.  In a quark gluon plasma
loop corrections are only negligible at very high temperatures due to
asymptotic freedom. Hence one should keep in mind that the leading order
result is in QCD only a valid approximation at high temperatures.

\subsection{Zero frequency limit}
Let us first rederive the zero frequency, zero momentum limit of the
Chiral Magnetic conductivity.  Since
\begin{equation}
 \lim_{p \rightarrow 0}
\lim_{p_0 \rightarrow 0}
\sum_{t = \pm} 
\log \left[
\frac{ (p_0 + i \epsilon + t q)^2 - (q + p)^2}
{ (p_0 + i \epsilon + t q)^2 - (q - p)^2}
\right] = \frac{2 p }{ q},
\end{equation}
it immediately follows from Eq.~(\ref{eq:GER5}) that
\begin{equation}
 \lim_{\vec p \rightarrow 0}
G^i_{\mathrm{R}}(p_0 = 0, \vec p)
= \frac{i e^2 \mu_5}{2 \pi^2} p^i.  
\end{equation}
As a result we recover the known zero frequency result of the Chiral
Magnetic conductivity for homogeneous magnetic fields ($\vec p
\rightarrow 0$),
\begin{equation}
 \sigma_0 \equiv  \sigma_{\chi}(\omega = 0)
= \frac{e^2}{2 \pi^2}  \mu_5.
\label{eq:rechicono0}
\end{equation}
The zero frequency limit is independent of $\mu$ and $T$.  Since this
value is constrained by the axial anomaly, loop corrections by gluons
and/or photons will not alter this result \cite{FKW}.

\subsection{Imaginary part}
To obtain the imaginary part of the Chiral Magnetic conductivity we
need to compute the imaginary part of the logarithm in
Eq.~(\ref{eq:GER5}). This is a sum of step functions times $\pi$.  We
find using that $q \geq 0$ and $p = \vert \vec p \vert \geq 0$,
\begin{multline}
\mathrm{Im} 
\sum_{t=\pm}(2q + t p_0)
\log \left[
\frac{ (p_0 + i \epsilon + t q)^2 - (q + p)^2}
{ (p_0 + i \epsilon + t q)^2 - (q - p)^2}
\right]
= \\
\pi \left[ 2 q - \vert p_0\vert \theta(p_0^2 - p^2) 
\right]
\left[\theta(q_+ - q)
-
\theta(q_- - q) \right]
\\
+ \pi p_0 \theta(p^2 - p_0^2)
\left[\theta(q - q_+)
+
\theta(q - q_-) \right]
,
\end{multline}
where $q_\pm = \tfrac{1}{2} \vert p_0 \pm p \vert$.

To compute the imaginary part of the Chiral Magnetic conductivity we
furthermore need the to perform the following two indefinite integrals
\begin{multline}
\int \mathrm{d} q\,
f(q) 
=
T \sum_{s, t=\pm}
s t \log \left[
1 + e^{ (q + t \mu_s) / T}
\right],
\\
\int \mathrm{d} q\,
 q f(q)
= \sum_{s, t=\pm} s t
\left \{
q T \log \left[
1 + e^{ (q + t \mu_s) / T}
\right]
\right.
\\
\left.
+ T^2 \mathrm{Li}_2 \left[
-e^{(q+ t \mu_s)/T}
\right]
\right \},
\end{multline}
here $\mathrm{Li}_2(z) = \sum_{k=1}^{\infty} z^k / k^2$ denotes a
polylogarithm of second order. We now find for the imaginary part of
the Chiral Magnetic conductivity at nonzero frequency and momentum
\begin{multline}
\sigma''_\chi(p)
=
\frac{e^2}{16 \pi} 
\frac{p^2\! -\! p_0^2}{p^3}
\Bigl \{
8 \theta(p^2 - p_0^2) 
 p_0 \mu_5 
\\
+ T \!\!\!
\sum_{s,t, r= \pm}
\!\!\!\!
s t 
\Bigl \{ 
2 r T \, \mathrm{Li}_2 \left[
-e^{(q_r + t \mu_s)/T}
\right]
\\
+
p \left[
\mathrm{sgn}(p_0) 
\theta(p_0^2 - p^2)
+ r \,
\theta(p^2 - p_0^2)
\right] \times
\\
\log \left[
1 + e^{ (q_r + t \mu_s) / T}
\right]
\Bigr \}
\Bigr \}.
\end{multline}

In the limit of homogeneous magnetic fields ($\vec p \rightarrow 0$),
the imaginary part of the Chiral Magnetic conductivity becomes
\begin{multline}
\sigma''_\chi(\omega)
=
\frac{e^2}{3 \pi} 
\omega \delta(\omega)
\mu_5
\\
+
\frac{e^2 \omega \vert \omega \vert}{96 \pi} 
\sum_{s,t= \pm} st
\left[
\frac{\mathrm{d} }
{\mathrm{d} q }
\tilde n(q + t \mu_s)
\right]_{q = \vert \omega \vert /2 }.
\label{eq:imchicon}
\end{multline}
The derivative with respect to the Fermi-Dirac distribution shows that
for small temperatures only states near the Fermi-surface contribute
to the Chiral Magnetic conductivity.

At zero temperature the imaginary part of the leading order
Chiral Magnetic conductivity becomes
\begin{equation}
\sigma''_\chi(\omega)
=
\frac{e^2}{3 \pi} 
\omega \delta(\omega)
\mu_5
-
\frac{e^2 \omega^2}{96 \pi} 
\sum_{s,t= \pm} st \,
\delta  (\omega / 2  + t \mu_s).
\label{eq:imagcondp0}
\end{equation}
The last equation shows that in the limit of $p \rightarrow 0$ the
leading order contribution to the Chiral Magnetic conductivity
develops resonances at $\omega = \pm 2 \mu_s$. As argued in the
previous section, these resonances can be attributed to
particle-antiparticle pair production in the time dependent
electromagnetic field.

For large temperatures $(T >\mu_5)$ we can approximate
Eq.~(\ref{eq:imchicon}) by
\begin{multline}
\sigma''_\chi(\omega)
=
\frac{e^2}{3 \pi} 
\omega \delta(\omega)
\mu_5
+
\frac{e^2 \omega \vert \omega \vert}{24 \pi T^2}
\tilde n( \vert \omega \vert / 2)^3 
\\
\times 
\left[
e^{\vert \omega \vert / T}
-
e^{\vert \omega \vert / (2 T)}
\right] \mu_5.
\end{multline}
The imaginary part at large temperatures has a maximum at $\omega / T
\approx 5.406$ with value $\sigma'' \approx 0.394 \sigma_0$.

\begin{figure}[t]
\includegraphics{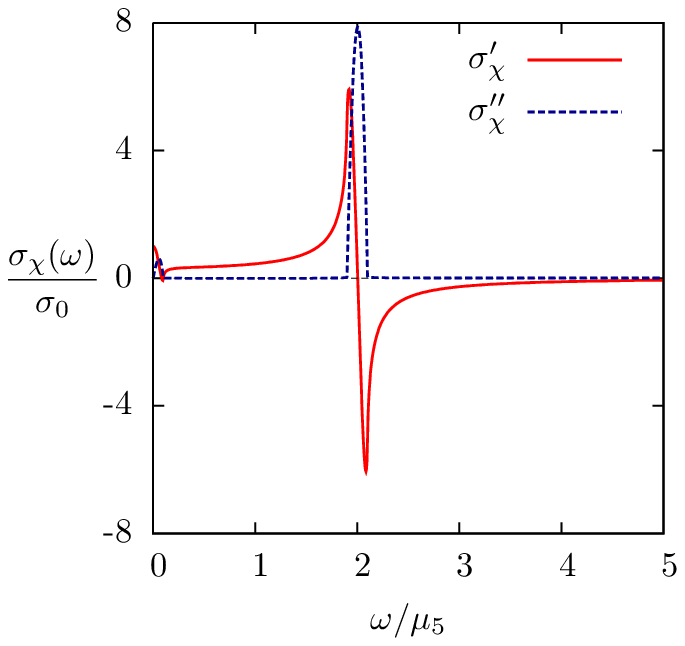}
\caption{Real (red, solid) and imaginary (blue, dashed) part of the
  leading order Chiral Magnetic conductivity as a function of frequency, 
  at $T=0$ and $\mu=0$ for $p=0.1\mu_5$. The result is normalized to the zero
  frequency conductivity $\sigma_0 = e^2 \mu_5 / (2 \pi^2)$. 
  \label{fig:sigmaT0}}
\end{figure}

\begin{figure}[t]
\includegraphics{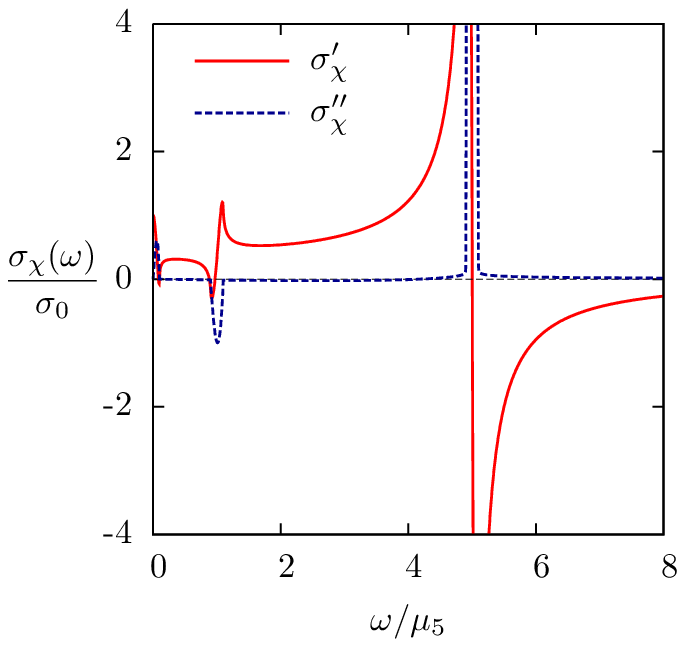}
\caption{Real (red, solid) and imaginary (blue, dashed) part of the
  leading order normalized Chiral Magnetic conductivity as a function
  of frequency, at $T=0$, $\mu=1.5\mu_5$ and $p=0.1\mu_5$.  The result
  is normalized to the zero frequency conductivity $\sigma_0 = e^2
  \mu_5 / (2 \pi^2)$.}
\label{fig:condT0mu}
\end{figure}

\begin{figure}[t]
\includegraphics{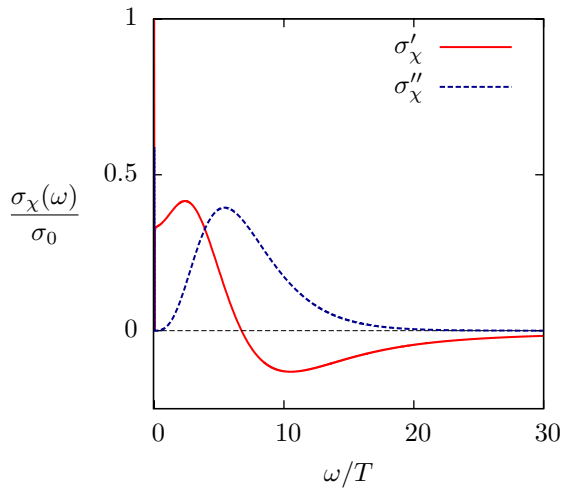}
\caption{Real (red, solid) and imaginary (blue, dashed) part of the
  leading order normalized Chiral Magnetic conductivity
at high temperatures ($T > \mu_5$) for homogeneous magnetic
fields ($p = 0$). At $\omega=0$ the normalized conductivity is equal
to $1$.
\label{fig:condtemp}}
\end{figure}

\subsection{Real part}
At zero temperature we can now recover the real part of the Chiral
Magnetic conductivity by applying the Kramers-Kronig relation,
Eq.~(\ref{eq:kk1}). We find
\begin{eqnarray}
\sigma'_\chi(\omega = 0) 
&=& \frac{e^2}{2\pi^2} \mu_5,
\\
\sigma'_\chi(\omega \neq 0)
&=&
\frac{e^2}{3 \pi^2}
\sum_{s=\pm} \frac{s \mu_s}{4 - (\omega / \mu_s)^2}.
\end{eqnarray}
For finite temperatures we were unfortunately not able to obtain
an analytic expression for the real part.

The delta function of the imaginary part at $\omega = 0$ only
contributes to $2/3$ of the real part at $\omega = 0$.  The other
$1/3$ part comes from the pair-production processes.  This conclusion
holds for any temperature. This is because the both the real (see
Eq.~(\ref{eq:rechicono0})) and imaginary part
(Eq.~(\ref{eq:imchicon})) at $\omega = 0$ are independent of
temperature. From the Kramers-Kronig relation it follows that for any
nonzero frequency the imaginary part at $\omega=0$ does not
contribute to the real part at $\omega \neq 0$. Therefore for any
temperature the real part of the Chiral Magnetic conductivity drops
from $\sigma_0$ at $\omega = 0$ to $\sigma_0 / 3$
just away from $\omega = 0$.

\subsection{Discussion}

We display the real and imaginary part for $T=0$, $p=0.1\mu_5$ and
$\mu=0$ in Fig.~\ref{fig:sigmaT0}. As was argued at the end of the
previous subsection, it can be seen in this figure that the real part
of the Chiral Magnetic conductivity drops from $\sigma_0$ at $\omega =
0$ to $\sigma_0 / 3$ just away from $\omega = 0$. Also the resonance
at $\omega = 2 \mu_5$ is clearly visible. The width of the imaginary
part at the resonance is equal to $2 p$. The real part of the
conductivity becomes negative above the resonance frequency. This is a
typical resonance behavior and implies that when the imaginary part
vanishes the response is $180$ degrees out of phase with the applied
magnetic field.

In Fig.~\ref{fig:condT0mu} we display the real and imaginary
part for $T=0$, $p=0.1\mu_5$ and $\mu=1.5 \mu_5$. In this case there
are resonances at $\omega = 5 \mu_5$ and $\omega = \mu_5$.  Equation
(\ref{eq:imagcondp0}) shows that the imaginary part is proportional to
$\omega^2$, therefore the second resonance at $\omega = 5 \mu_5$ is
much stronger than the first one at $\omega = \mu_5$. Because the
second resonance is due to the right-handed modes, and the first one
due to left-handed, the contribution of the second resonance has
opposite sign to the first resonance.

The real and imaginary part of the Chiral Magnetic conductivity at
high temperatures ($T>\mu_5$) are displayed in
Fig.~\ref{fig:condtemp}. This figure is the most relevant for QCD at
very high temperatures, since then loop corrections will be small. As
argued in the previous subsection it can be seen in the figure that
the real part of the conductivity drops from $\sigma_0$ at $\omega=0$
to $\sigma_0/3$ just away from $\omega=0$.

Let us now study the induced current in a magnetic field of the form
created during heavy ion collisions.  For simplicity we approximate
the two colliding nuclei by point like particles like in
Ref.~\cite{MM96}. This gives a reasonable approximation to the more
accurate methods discussed in Refs.~\cite{KMW, SIT09} and is most
reliable for large impact parameters. The magnetic field at the center
of the collision can then be written as
\begin{equation}
B(t) = \frac{1}{\left[1+ (t/\tau)^2\right]^{3/2}} B_0,
\label{eq:magfield}
\end{equation}
with $\tau = b / (2\sinh Y)$ and $e B_0 = 8 Z \alpha_{\mathrm{EM}}
\sinh Y / b^2$.  Here $b$ denotes the impact parameter, $Z$ the charge
of the nucleus, and $Y$ the beam rapidity. For Gold-Gold ($Z=79$)
collisions at 100 GeV per nucleon one has $Y=5.36$. At typical large
impact parameters (say $b=10\;\mathrm{fm}$) one finds $e B_0 \sim 1.9
\times 10^5 \;\mathrm{MeV^2}$ and $\tau = 0.05\; \mathrm{fm}/c$. For
31 GeV per nucleon ($Y=4.19$) Gold-Gold collisions one finds at
$b=10\;\mathrm{fm}$, $e B_0 \sim 5.9 \times 10^4 \;\mathrm{MeV^2}$ and
$\tau = 0.15\;\mathrm{fm}/c$.  The Fourier transform of
Eq.~(\ref{eq:magfield}) equals
\begin{equation}
\tilde B(\omega) 
= 2 \tau^2 \vert \omega \vert K_1(\tau \vert \omega \vert) B_0,
\label{eq:fouriermagfield}
\end{equation}
where $K_1(z)$ denotes the first-order modified Bessel function of the second
kind.

\begin{figure}[t]
\includegraphics{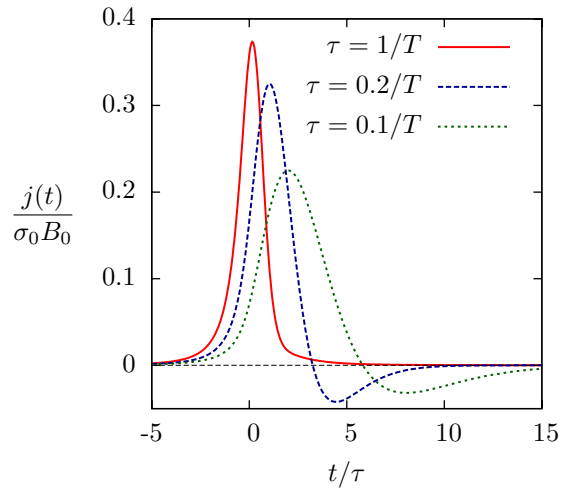}
\caption{Induced current in time-dependent magnetic field,
  Eq.~(\ref{eq:magfield}), as a function of time, at very high temperature. 
  The results are plotted for different values of the characteristic time
  scale $\tau$ of the magnetic field. 
  \label{fig:jz}}
\end{figure}

For illustration purposes we will assume that our magnetic field is
(unlike in heavy ion collisions) homogeneous. The induced current can
be found by applying Eq.~(\ref{eq:response}). We display the induced
current in the magnetic field of Eq.~(\ref{eq:magfield}) in a system
with nonzero chirality at very high temperatures in Fig.~\ref{fig:jz}.
The induced current is plotted as a function of time for three
different characteristic time scales $\tau$ of the magnetic field.

In any general decaying magnetic field, the only relevant frequencies
are the ones which are smaller than of order the inverse life-time of
the magnetic field, $\omega \lesssim \mathcal{O}(1/\tau)$. In
Fig.~\ref{fig:condtemp} it can be seen that even in the
non-interacting case there still is sizable response at high
temperatures as long as $\tau \gtrsim 1 / (5 T)$. Hence even for such
fast changing fields there will really be a induced current, although
it will be about $1/3$ of the adiabatic approximation which is $j(t) =
\sigma_0 B(t)$. This can be more clearly seen in Fig.~\ref{fig:jz}.
The red solid line describes a relatively slowly varying field with
$\tau = 1/T$, in that case indeed $j(t) \sim \sigma_0 B(t) / 3$.  The
red solid line can therefore also be seen as a guide to eye of the
time dependence of the magnetic field. Realizing this, in a fast
changing magnetic field with $\tau=0.1/T$ it clearly takes some time
for the current to respond. The maximal current will be smaller, but
it takes also more time for the current to diminish after the magnetic
field has gone away. At late times the current can even become
negative.  The situation in which $\tau=0.1/T$ approximately
corresponds to 100 GeV per nucleon Gold-Gold collisions at
$b=10\;\mathrm{fm}$.

Thermal fluctuations will increase in magnitude when the temperature
is increased.  These fluctuations can cause the spins of the particles
to align along the magnetic field.  Hence one would expect that when
keeping $\tau$ fixed and increasing the temperature the system will
respond faster to the changing magnetic field.  This can be seen in
Fig.~\ref{fig:jz}, by increasing the temperature at fixed $\tau$ one
goes from the dotted to the solid line, which indicates faster
response.

The discussion will alter when loop corrections due to interactions
are taken into account. In that case the quasi-particles obtain a
thermal width $\Gamma$ which is of order $\alpha T$. Because of the
thermal broadening we expect that the peak in the Chiral Magnetic
conductivity (Fig.~\ref{fig:condtemp}) at $\omega = 0$ will get a
width of order $\Gamma$. At the same time the value at $\omega = 0$
will not change since it is constrained by the anomaly.  As long as
$\tau \gtrsim \mathcal{O}(1/\Gamma)$ the zero frequency result will
therefore be a reasonable estimate and the induced current will be
more or less equal to the adiabatic approximation $j(t) \sim \sigma_0
B(t)$.  Hence the stronger the system interacts at fixed $\mu_5$ and
$T$, the stronger the response at small frequencies.  Qualitatively
one would expect such behavior, because the stronger a particle
interacts the faster the particle aligns its spin and hence momentum
along the field.

Let us finally discuss how much charge is separated by the current.
Let us divide our infinite system into an upper half ($u$) and a lower
half ($l$), and set the boundary to be the plane $z=0$. Using the the
fact that the current is conserved $\partial_\mu j^\mu = 0$ one finds
that the change in charge in the upper hemisphere per unit of time
equals
\begin{equation}
  \frac{\mathrm{d} Q_u}{\mathrm{d} t} = 
  - \int_u \mathrm{d}^3 x \, \vec \nabla
  \cdot \vec j = \int \mathrm{d}^2 x \,j_z(z=0,t).
\end{equation}
By integrating this equation over time and using
Eq.~(\ref{eq:response}) we find that the change in charge in the upper
hemisphere $\Delta Q_u$ equals
\begin{equation}
\Delta Q_u = A \int_{-\infty}^{\infty} \mathrm{d} t\, j(t)
= \sigma_0 A \tilde B(\omega = 0),
\label{eq:deltaqu}
\end{equation}
where $A$ is the area of the plane $z=0$. Because of global charge
conservation the charge change in the lower hemisphere is equal to
$\Delta Q_l = -\Delta Q_u$.  In Eq.~(\ref{eq:deltaqu}) we find the
surprising result that as long as linear response is valid and the
chirality is constant the total induced charge difference between
upper and lower hemisphere is independent of detailed dynamics and
just determined by the zero frequency chiral magnetic conductivity
$\sigma_0$. In the magnetic field of Eq.~(\ref{eq:magfield}) we obtain
for the induced charge difference between the upper and lower
hemisphere
\begin{equation}
\Delta Q \equiv \Delta Q_u - \Delta Q_l = 4 \sigma_0 A \tau B_0. 
\end{equation}

\section{Conclusions}
In a quark gluon plasma it is possible to generate nonzero chirality
by gluon configurations with nonzero topological charge.  In the
presence of a magnetic field nonzero chirality leads to a current
along the field. This is the Chiral Magnetic Effect which can
potentially give rise to observable effects in heavy ion collisions.
Since the magnetic field in heavy ion collisions is rapidly decreasing
as a function of time, it is desirable to study the Chiral Magnetic
Effect in a time-dependent magnetic field. In this article, we have
shown for the first time that such study is possible in a systematic
way using linear response theory.

To obtain the induced current in a time-dependent magnetic field we
have derived a general Kubo-formula for the Chiral Magnetic
conductivity. We have shown that the Chiral Magnetic conductivity is
proportional to the antisymmetric part of the off-diagonal photon
polarization tensor.

Since we have applied linear response theory, our results are only
valid for small magnetic fields. This means that the magnetic fields
should not alter the plasma dynamics much, implying that the magnetic
length $1/\sqrt{eB}$ has to be larger than the (color) electric screening
length $\sim 1/g T$.  Hence for reliable results the magnetic
field should satisfy $eB \lesssim g^2 T^2$.

We have computed the leading order Chiral Magnetic conductivity for
constant chirality using perturbation theory. As such our result is
only applicable for QCD at high temperatures where loop corrections
can be neglected. We have shown that pair production in the
time-dependent electromagnetic field gives rise to non-trivial
behavior of the Chiral Magnetic conductivity at nonzero frequencies.
Our result can be systematically improved by including loop
corrections, in a similar way to what has been done for the electrical
conductivity in for example Ref.~\cite{AR}.

The general formula for the Chiral Magnetic conductivity we have
obtained allows for the evaluation using other methods.  For example
as suggested in Ref.~\cite{RSS09} one could study the Chiral Magnetic
Effect in holographic models of QCD. Also since $\mu_5$ does not give
rise to a sign problem \cite{FKW}, one could in principle study the
Chiral Magnetic conductivity using lattice QCD (see also
Ref.~\cite{BCLP09} for another approach). However, because one has to
perform an analytic continuation in order to obtain the retarded
correlator this is not completely straightforward to do.

The main message of this calculation is the finding that even in the
leading order the Chiral Magnetic conductivity has sizable response at
nonzero frequencies. This is a clear proof that even a rapidly 
decaying magnetic can give rise to a non-negligible current.
By taking into account interactions the quasi-particles will obtain
a thermal width, which as we argued will increase the response at
small frequencies but does not change the zero frequency result.

This calculation of the Chiral Magnetic conductivity is a small step
in order to improve our understanding of the dynamics of the Chiral
Magnetic Effect.  To apply our results to heavy ion phenomenology one
has to take into account also other dynamical effects like the
time-dependence of the chirality, the radial flow, and possible
screening mechanisms. Also it would be interesting to study the
effects of a time-dependent magnetic field on other physical
quantities and effects, like for example the chiral condensate, the
chiral phase transition and dynamical chiral symmetry breaking. So far
these have only been investigated in a constant magnetic field
\cite{GMS, SS97,CGW07, AF08, SA08, FM08, BC09}.

\section{Acknowledgments}
We are grateful to Kenji Fukushima, Larry McLerran, Dirk Rischke and
Andreas Schmitt for discussions.  This manuscript has been authored
under Contract No.~\#DE-AC02-98CH10886 with the U.S.\ Department of
Energy. The work of H.J.W.\ was supported partly by the Alexander von
Humboldt Foundation and partly by the ExtreMe Matter Institute EMMI in
the framework of the Helmholtz Alliance Program of the Helmholtz
Association (HA216/EMMI).

\end{document}